\documentclass[twocolumn]{aastex631}

\usepackage{makecell}

\shorttitle{Barred galaxy at z$=$3.16}
\shortauthors{Cheng et al.}

\begin{document}

\title{Discovery of a Barred-Spiral Galaxy at $z_{\rm spec}=3.16$ II. The Star Formation History}

\author[0000-0001-8551-071X]{Yingjie Cheng}
\affiliation{Department of Astronomy, University of Washington, Seattle, WA 98195, USA}
\affiliation{University of Massachusetts Amherst, 710 North Pleasant Street, Amherst, MA 01003-9305, USA}
\email{yingjiec@uw.edu}

\author[0000-0002-7831-8751]{Mauro Giavalisco}
\affiliation{University of Massachusetts Amherst, 710 North Pleasant Street, Amherst, MA 01003-9305, USA}
\email{mauro@umass.edu}

\author[0009-0005-7495-3298]{Daniel Ivanov}
\affiliation{Department of Physics and Astronomy, University of Pittsburgh, Pittsburgh, Pennsylvania 15260, USA}
\affiliation{University of Massachusetts Amherst, 710 North Pleasant Street, Amherst, MA 01003-9305, USA}
\email{DAI34@pitt.edu}

\author[0000-0001-6820-0015]{Luca Costantin}
\affiliation{Centro de Astrobiolog\'ia (CAB), CSIC-INTA, Ctra de Ajalvir km 4, Torrej\'on de Ardoz, 28850, Madrid, Spain}

\author[0000-0003-2676-8344]{Elena D'Onghia}
\affiliation{Department of Astronomy, University of Wisconsin, Madison, WI 53706, USA}

\author[0000-0002-4162-6523]{Yuchen Guo}
\affiliation{Department of Astronomy, The University of Texas at Austin, Austin, TX, USA}

\author[0000-0002-1590-0568]{Shardha Jogee}
\affiliation{Department of Astronomy, The University of Texas at Austin, Austin, TX, USA}

\author[0000-0003-1614-196X]{John R. Weaver}\thanks{Brinson Prize Fellow}
\affiliation{MIT Kavli Institute for Astrophysics and Space Research, 70 Vassar Street, Cambridge, MA 02139, USA}
\email{john.weaver.astro@gmail.com}

\author[0000-0001-7160-3632]{Katherine E. Whitaker}
\affiliation{University of Massachusetts Amherst, 710 North Pleasant Street, Amherst, MA 01003-9305, USA}

\begin{abstract}

We present a detailed analysis of a massive barred galaxy at $z_{\rm spec}=3.1591$ using deep multi-band imaging from HST and JWST. For the first time, we resolve its morphology and stellar structures thanks to the JWST/NIRCam NIR and MIR photometry. The galaxy possesses two distinct components with significantly different colors. Through careful image decomposition and masking, we isolate and characterize the flux contribution from each component. The galaxy exhibits a clear spiral morphology, and in a separate companion paper, we present evidence suggesting the presence of a stellar bar. Based on spatially resolved spectral energy distribution modeling with \texttt{Prospector}, we derive the star formation history and other physical properties of the bar and the surrounding regions. The total stellar mass of the galaxy is constrained as $\log(M_*/M_{\odot}) = 10.63\pm0.13$. We find that the bar region contains around 30\% of the total stellar mass, but only accounts for around 8\% of the recent star formation rate. The region containing the potential bar shows a significantly older mass-weighted stellar age, supporting the inside-out scenario for galaxy formation, and providing tentative evidence for bar quenching in the early stage. The quick onset of a stellar bar at this redshift requires a low dark matter fraction, suggesting the baryon-dominated nature of high-$z$ massive galaxies, and offering rare insight into galaxy evolution at around two billion years after the Big Bang.

\end{abstract}

\keywords{High-redshift galaxies(734) --- Galaxy bars(2364) --- Spectral energy distribution(2129)}

\section{Introduction} \label{sec:intro}

Disk-like galaxies dominate the galaxy population in our local universe. Around 65\% of galaxies host well-developed stellar disks from redshift 0 up to $z\sim$2 (e.g., \citealt{2011ApJ...730...38V, 2014ApJ...792L...6V}). In those galaxies, galactic disks are key regions for star formation as a result of the high concentration of gas and dust. In recent years, thanks to its high sensitivity and resolution in the infrared, the James Webb Space Telescope (JWST) has updated our knowledge of the traditional Hubble sequence \citep{2023ApJ...942L..42R, 2023ApJ...948L..13J}, especially at high redshifts. The fraction of disk-like galaxies does not decline as rapidly as expected with increasing redshift, and they are still prevalent up to redshifts $z\sim$6 (e.g., \citealt{2023ApJ...946L..15K, 2023ApJ...955...94F, 2025A&A...699A.360C}). Well-developed disks (both in stellar and gaseous phases) at early cosmic times have become one of the most pressing problems in galaxy formation and evolution studies.

Disk-like galaxies often suffer from gravitational instabilities that cause the formation of bars and spiral arms \citep{1977ARA&A..15..437T, 2004ARA&A..42..603K}. Stellar bars play a crucial role in the secular evolution of disk galaxies in terms of both star formation and multi-phase dynamics, and thus can be a potential tracer of the buildup of disks and their host dark matter halos \citep{2006ApJ...645..209D, 2014RvMP...86....1S}. The fraction of barred galaxies relative to the total number of disk-like galaxies, or the bar fraction, provides a valuable test for theories of disk evolution. According to early studies, the bar fraction is $\sim$60\% in the local universe, and declines with increasing redshift up to $z\sim$1 (e.g., \citealt{2004ApJ...612..191E, 2007ApJ...657..790M, 2007ApJ...659.1176M, 2008ApJ...675.1141S}). Measuring the bar fraction at higher redshifts can be challenging due to the decreasing bar sizes and observational effects that impact the visibility of small-scale features \citep{2025MNRAS.tmp.1892L, 2023ApJ...945L..10G, 2025ApJ...985..181G}. 

Recently, with the unprecedented angular resolution and wavelength coverage of JWST, observations at $z>2$ have witnessed a wide diversity of galaxy morphologies, with a large fraction showing disk-like or peculiar features (e.g., \citet{2023ApJ...946L..15K, 2023ApJ...948L..13J, 2024A&A...685A..48H}). An important conclusion is that galaxies with well-established disks and/or bars exist across a long cosmic time, and further high-$z$ observations are needed to specify when these features first formed. In particular, well-developed bars are found as early as z$\sim$3 in the stellar density field of some disk galaxies (e.g., \citet{2023Natur.623..499C, 2023ApJ...945L..10G, 2025ApJ...985..181G}). The bar fraction is now constrained up to $z\sim$3.5, with a value of around 5\% for massive galaxies with $\log (M_*/M_{\odot})>10$ \citep{2025ApJ...985..181G, 2025MNRAS.tmp.1892L}. How to form disks and bars in the first few billion years after the Big Bang is an open question in modern galaxy evolution theories.

Numerical simulations are now probing the growth of bars out to redshift 4 (e.g., TNG50, \citet{2022MNRAS.512.5339R, 2022ApJ...934...52B, 2025MNRAS.538.1587F}), although the impact of feedback, turbulence, and tidal interactions is highly uncertain. The bursty and chaotic stellar density field at early times adds to the complexity of modeling. According to 3D hydrodynamic simulations \citep{2023ApJ...947...80B, 2024ApJ...968...86B}, the timescale of bar formation is sensitive to the disk-to-total mass ratio ($f_{disk}$) within the disk scale length (2.2$R_{disk}$). At $f_{disk}>$0.3, the bar formation time declines exponentially with increasing $f_{disk}$, leading to a high bar fraction at high redshifts. Meanwhile, a strong negative trend in the dark-matter fraction with increasing redshift is reported, where the inner baryons dominate over dark matter in early massive galaxies, and the effect appears to increase with redshift  (e.g., \citet{2017Natur.543..397G}). Therefore, a large fraction of massive, high-redshift galaxies were strongly baryon-dominated, setting the stage for the quick onset of stellar bars. Confirming the existence of bars at ultra-high redshifts and studying the properties of their host galaxies will further constrain the mechanism of bar formation, and also provide information on the dark matter fraction in the early universe.

Bars have a great impact on galaxy evolution through driving gas inflows, helping build bulges, and redistributing matter and angular momentum within the galaxies. In the local universe, bars are found to enhance star formation in their central region \citep{2004ARA&A..42..603K, 2005ApJ...630..837J} and beyond their ends, while suppressing star formation in their arms \citep{2024ApJ...973..129G}. This might suggest an inside-out formation and quenching scenario, where bars can trigger a starburst in the center of the galaxy after causing a substantial inflow of gas, which accelerates the depletion of gas reservoirs and eventually quenches the host galaxy. Due to limited angular resolution, spatially resolved studies of bars in galaxies are rare at high redshift. Investigating the spatially resolved properties and star formation histories of barred galaxies at $z>2$, both inside and outside of their bar structures,  can provide valuable insights into quenching mechanisms in the early universe and allow for meaningful comparisons with those observed in the local universe.

The identification of stellar bars in spiral galaxies up to $z\approx 3$ with JWST imaging \citep{2023Natur.623..499C, 2023ApJ...945L..10G, 2024MNRAS.530.1984L, 2025ApJ...985..181G, 2025ApJ...987...74G} has opened up the possibility of investigating the role played by bars in the bulge formation and Active Galactic Nucleus (AGN) growth. In addition, the mechanisms and conditions that cause instabilities leading to bar formation and the associated time scales can now also be probed. Here we report on the properties of the stellar populations of a massive barred spiral galaxy at $z>3$, a mere $\approx 2$ Gyr after the Big Bang, which was discovered in the JWST/NIRCam multi-band imaging survey of the COSMOS-PRIMER field \citep{2021jwst.prop.1837D}. The galaxy was first identified as a Lyman-break galaxy (LBG) at $z_{\rm spec}=3.1591$ as part of the MOSFIRE Deep Evolution Field (MOSDEF) survey \citep{2015ApJS..218...15K}, while NIRCam imaging at wavelengths redder than its Balmer break reveals further details. At these wavelengths, the bulge-bar complex and the spiral structure become visible and we report on the identification of the bar and its size in a companion paper (Ivanov et al. in prep.). The rest of the paper is organized as follows. In Section \ref{sec:obs}, we describe the observed data, imaging analysis, and our SED modeling process. We present our main results in Section \ref{sec:res} and further discuss our findings in Section \ref{sec:dis}. The conclusions of this paper are summarized in Section \ref{sec:con}.

Throughout this paper, we assume a flat cosmology with $\Omega_M$ = 0.3, $\Omega_\Lambda$ = 0.7 and Hubble parameter $H_0 = 70 \textrm{ km}^{-1}\textrm{s}^{-1}\textrm{Mpc}^{-1}$. All the magnitudes are given in the AB system \citep{1983ApJ...266..713O}.

\section{Observations and Data Analysis} \label{sec:obs}

\begin{figure*}[ht!]
\epsscale{1.2}
\plotone{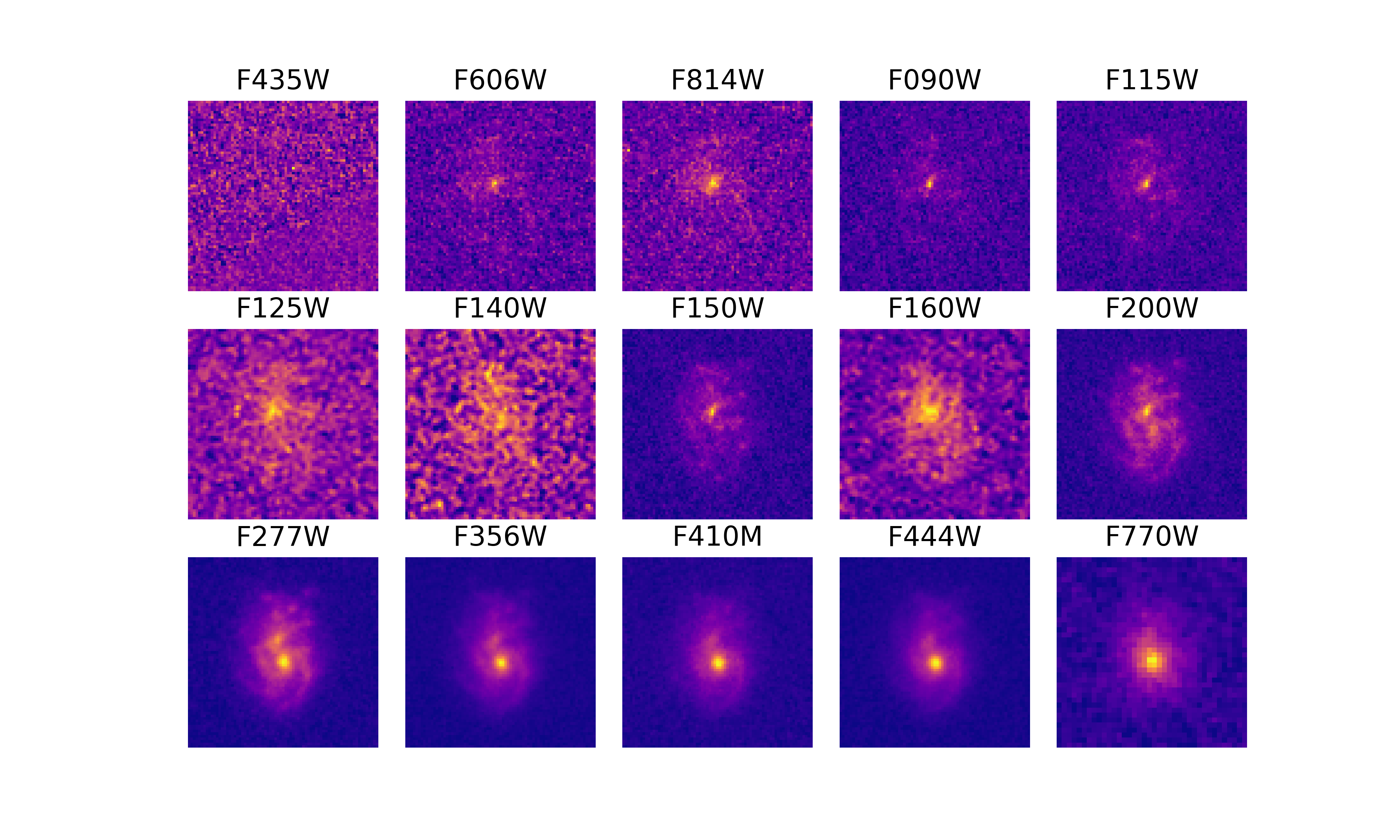}
\plotone{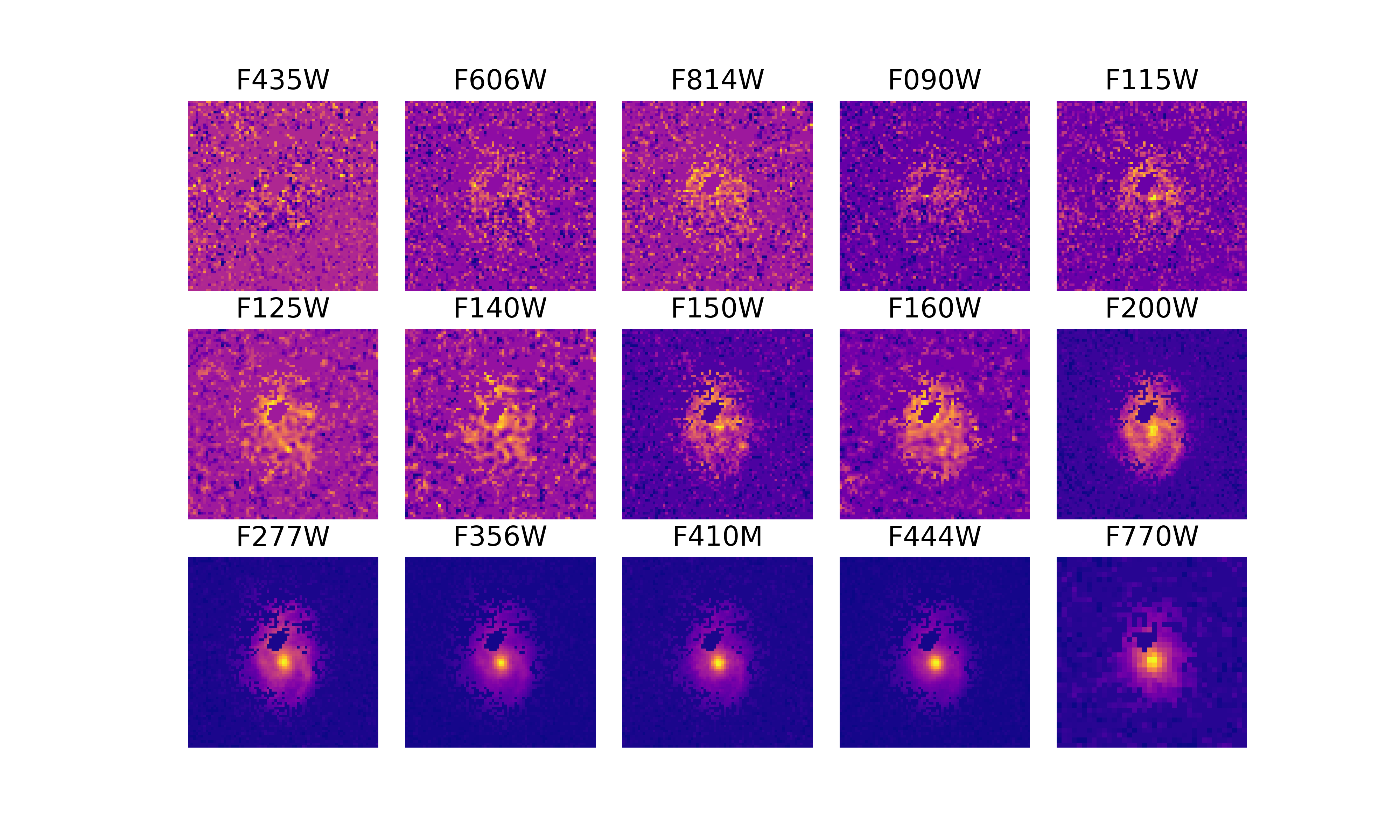}
\caption{The postage stamps of HST and JWST images for COSMOS-74706. The upper panel shows the original images, and the lower panel shows images after masking the blue upper blob.
\label{fig:stamp}}
\end{figure*}

\begin{figure*}[ht!]
\plottwo{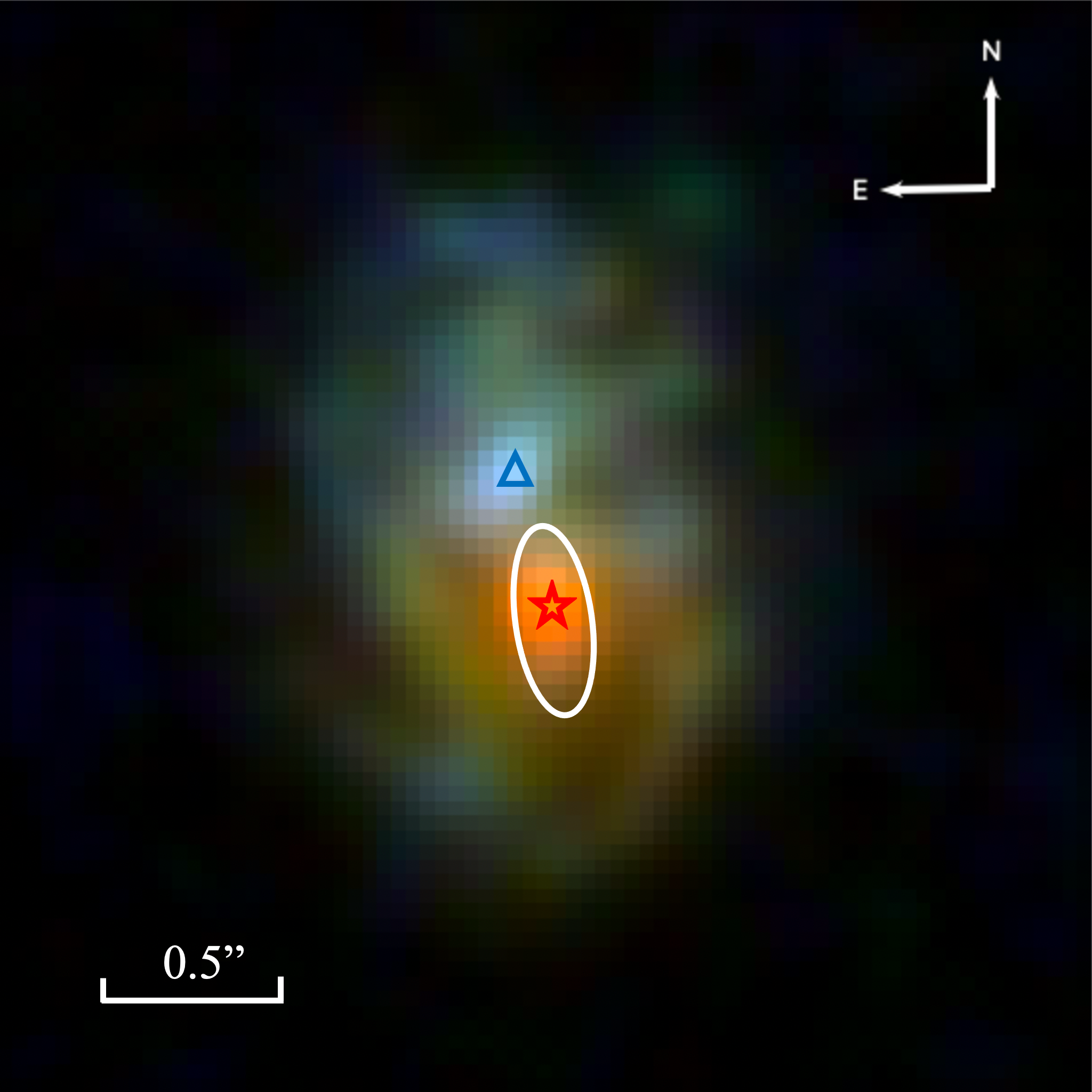}{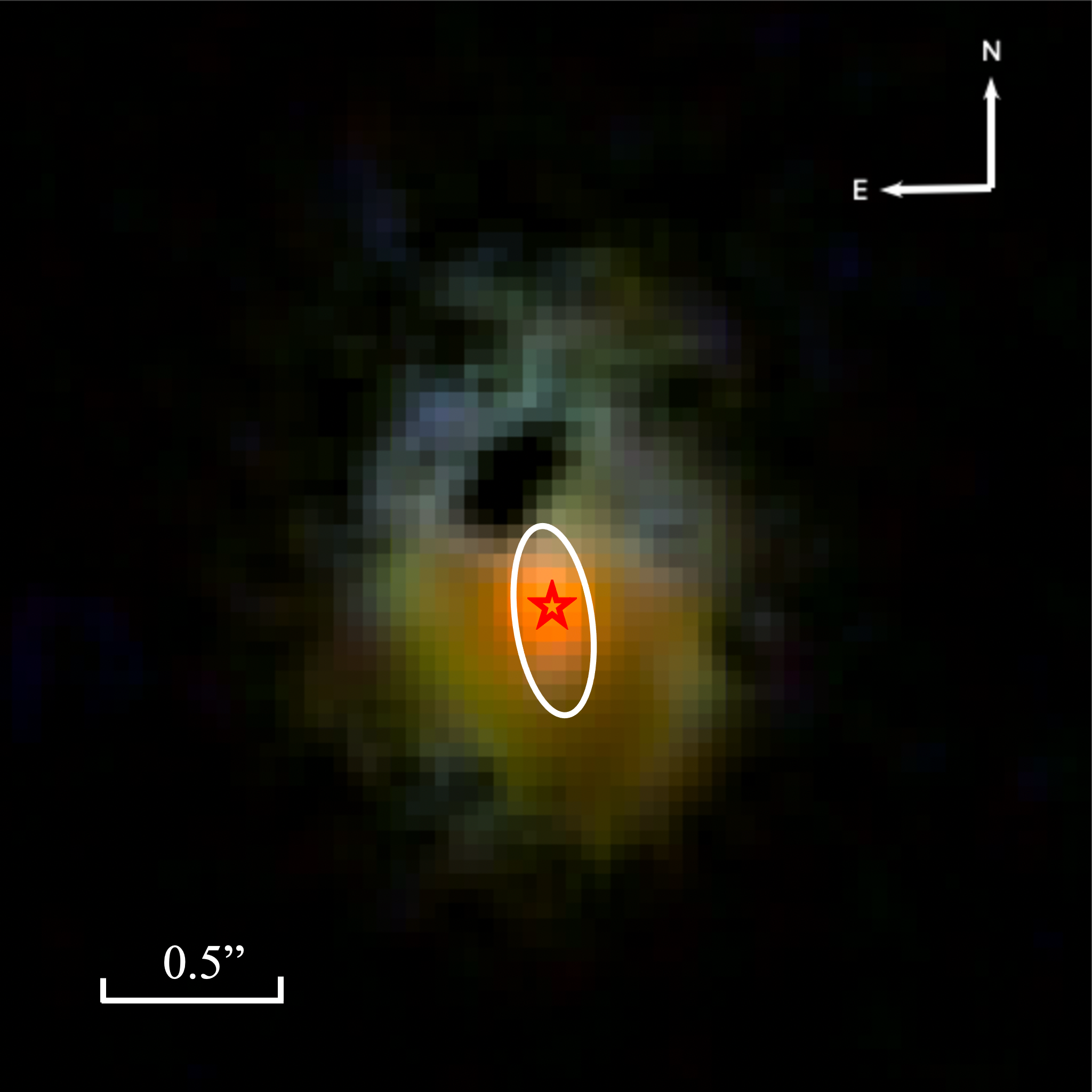}
\caption{Left: the PSF-matched RGB image of COSMOS-74706 with F356W in red, F200W in green, and F115W in blue. The upper blob is represented by a blue triangle, while the lower component is denoted by a red star. The visually identified bar region is shown within a white ellipse. Right: the same as the left panel, but after masking the blue component.
\label{fig:rgb}}
\end{figure*}

The spectroscopic data from Keck I MOSFIRE/MOSDEF \citep{2015ApJS..218...15K} show strong [OIII]5007 and H$\beta$4861 lines in the K band, and [OII]3729 line in the H band, anchoring the redshift of this galaxy precisely at $z=$3.1591.
In this study, we utilize data from the Public Release IMaging for Extragalactic Research survey (PRIMER, GO 1837, PI: James Dunlop). PRIMER is a deep, wide-area JWST imaging survey over the HST/CANDELS COSMOS \citep{2007ApJS..172....1S} and UDS \citep{2007MNRAS.379.1599L} fields. The NIRCam data are reduced from the v7 mosaics using GRIzLI\footnote{\url{https://github.com/gbrammer/grizli}}, and the catalogs are built with the aperture photometry code \textsc{aperpy}\footnote{\url{https://github.com/astrowhit/aperpy}} (see \citealt{2024ApJS..270....7W} for detailed settings). The source considered here is listed in the catalog as COSMOS-74706, which has J2000 sky coordinates RA$=$150.1007649 and DEC$=$2.341443311, coincident with source 3DHST-15810 (catalog version v4.1) with coordinates RA$=$150.100800, DEC$=$2.341475. We take all available data for it in the PRIMER internal data release V2.0.0, including JWST/NIRCam photometry in the F090W, F115W, F150W, F200W, F277W, F356W, F410M and F444W bands, JWST/MIRI photometry in the F770W band, and HST/WFC3 photometry in the F435W, F606W, F814W, F125W, F140W, and F160W bands.

The unique morphology and spectral energy distribution of COSMOS-74706 add to the mystery of this system. The overall morphology is consistent with that of a nearly face-on spiral galaxy. Through high-resolution JWST imaging, we visually recognize two distinct components of this system with vastly different colors: a blue upper blob and a red lower bulge-like component. Signs of bar features surrounding the red component have been identified by three independent methods: isophotal ellipse fitting, multicomponent GALFIT \citep{2010AJ....139.2097P} profile fitting, and Fourier multiplicity decomposition (see Ivanov et al. in prep for details). With the F200W image where the bar is most visible, the bar semi-major axis is 0.18$^{\prime\prime}$ or 1.4 kpc at z$=$3.1591. The rise in ellipticity over the bar region is around 0.2, and the bar modulus $B_M$ measured by Fourier analysis \citep{2017A&A...601A.132G} reaches up to 12.24 which is in excess of the threshold calibrated on a sample of deprojected $z\sim 1-3$ disk galaxies in Paper I.

Most of the spiral arms and bar features are largely invisible in HST images, especially for ACS filters at the bluer end. The bright blue blob identified in HST images (blue triangle shown in Fig.\ref{fig:rgb}) distinctly offsets the red component with barred-spiral features (red star in Fig.\ref{fig:rgb}) by $\sim$0.35$^{\prime\prime}$. Given the significantly different behaviors in photometry between the upper and lower blobs, it is highly likely that the system consists of two components with divergent physical properties. Although the emission lines detected in this system suggest a redshift of 3.1591 with high confidence \citep{2015ApJS..218...15K}, we do not rule out the possibility that the system is composed of two different galaxies that closely overlap with or interact with each other.

The upper blob is bright in all filters (except for the F435W image, which is too noisy) and aligns well with the emission line detections that indicate a redshift of 3.1591. The lower component is only revealed at longer wavelengths, possibly indicating heavier dust obscuration, more quenched star formation, or a background galaxy at a higher redshift. If another galaxy exists in the system, the visual Balmer break from the imaging suggests $z\gtrsim$5, while detailed SED modeling and/or further spectroscopic observations are needed to confirm its authenticity and the accurate redshift. Currently, no confirmed emission lines associated with a potential background galaxy have been detected. 
Alternatively, if the system is a single galaxy with large color gradients, the spatial variation of its multiple physical properties is yet to be explored. In this work, we will test and evaluate the proposed explanations for the bimodal structure. By masking the blue upper blob, we will take a close look at the morphology and SED of the red component, and in particular, its bar features. With spatially resolved SED modeling, we will compare the physical parameters and star formation histories within and outside the bar region, discussing the implications for galaxy evolution related to bar structures and more broadly, the disk instability.


\subsection{Decomposition and Masking}
\label{sec:mask}

To better analyze the properties of the inner red component and test the presence of a potential background galaxy, we need to remove the flux contribution from the blue upper blob. At redshift $z\sim$5.5, the 912-1216$\mathring{A}$ Ly$\alpha$ break shifts to 5928-7904$\mathring{A}$, and falls into the coverage of F606W and F814W bands of HST. Therefore, the red component and its surroundings are very faint in HST/ACS filters (F435W, F606W, F814W). This assumption is reinforced by Fig. \ref{fig:stamp}, where the lower part of the system is largely invisible in HST/ACS images. By stacking the HST/ACS images, we create an image dominated by the blue component, revealing its spatial coverage. Due to the low signal-to-noise ratio in the F435W band, we stack only the F606W and F814W images and then estimate the local background noise $\sigma$ with \texttt{Photutils}. All the images are PSF-matched to F444W before stacking and masking based on the PSF measurements from PRIMER internal data release.

We start by masking bright pixels in the stacked image to remove the contribution of the blue component. All pixels with flux values greater than a certain threshold (e.g., 3$\sigma$) in the stacked image are marked as the blue components. By masking those pixels from the original image in each available filter, we obtain a set of masked images. However, owing to the high overlap of the two components/galaxies, a single flux threshold cut is not optimized. Given the distinct difference between their colors, we test the validation of color selection with different combinations of filters. Pixels with rising SED at the blue end but which get fainter at the red end are more likely to belong to the blue component. Combined with visual inspections, our final selection criterion for the blue component is a combination of color cut and flux threshold cut:

$$
\left\{
    \begin{array}{l}
    F115W - F200W < 0~\textrm{nJy} \\
    F200W - F356W > -0.3~\textrm{nJy} \\
    F606W + F814W > 4\sigma \, (\sim1.78~\textrm{nJy})
    \end{array}
\right.
$$
\vspace{0.2cm}

Pixels that meet all three criteria are considered part of the blue upper component and are then masked from the original image. The final masked image of each band is shown in the lower panel of Fig. \ref{fig:stamp}. The barred-spiral features are clearly visible in the masked JWST/NIRCam images at F200W and beyond. We apply 2D interpolations on masked pixels using \texttt{scipy.interpolate} with Delaunay triangulation to enable further morphological analysis on the masked images. 
When performing the isophotal ellipse-fitting analysis and the Fourier decomposition analysis for the interpolated image, the presence of a stellar bar holds (see Ivanov et al. in prep. for details).

Given the masked images, we measure the multi-band fluxes through aperture photometry with a large aperture radius of 1.5$^{\prime\prime}$, and run SED modeling with \texttt{Prospector} \citep{2021ApJS..254...22J}. Using the same model settings, we observed an improvement in goodness-of-fit when employing masking. This suggests that separating the contributions from the upper and lower components simplifies the SED analysis of this system.

\subsection{Spatially Resolved SED Modeling}

We start by modeling the integrated photometry from the original images. Given all available photometry data from HST and JWST, we run SED fitting with \texttt{Prospector}, which adopts the stellar population synthesis model from FSPS \citep{2009ApJ...699..486C} and nebular emission model from \citealt{2017ApJ...840...44B}. We assume the \citet{2003PASP..115..763C} initial mass function (IMF) and the Calzetti dust attenuation law \citep{2000ApJ...533..682C} in all the fits. The parameter settings are detailed below.

The redshift is sampled with an initial guess of $z=3$ and a uniform prior between $z=2$ and $z=4$. The optical depth of the V band diffuse dust (the `dust2' parameter in FSPS) is sampled with a uniform prior between 0 and 4 and an initial guess of 1. The logarithm of stellar metallicity (the `logzsol' parameter in FSPS) is sampled with a uniform prior between -2 and 0.19, starting at $\log (Z/Z_{\odot})=0$ (solar metallicity). The logarithm of total stellar mass is sampled with a uniform prior from 8.5 to 12.5, starting at $\log (M_*/M_{\odot})=10$. Dust emission and nebular emission (line and continuum) are turned on in the fitting. We adopt a non-parametric SFH algorithm with 5 lookback time bins and the `continuity prior' to control the smoothness in SFR (see also \citealt{2019ApJ...876....3L, 2021ApJS..254...22J}). The first lookback time bin is fixed at $0<t<30\ \mathrm{Myr}$ to capture recent episodes of star formation, and the last bin is set to be $85\%-100\%$ of the age of the universe at the given redshift. All the intervening bins are evenly spaced on $log(t)$ scale. The sampling is performed with the dynamical nested sampling code $Dynesty$ \citep{2020MNRAS.493.3132S}.

The photometric redshift given by \texttt{Prospector} fit is $z=3.07^{+0.10}_{-0.13}$, which is consistent with the spectroscopic redshift found by Keck observations and disfavors the presence of a high-$z$ background galaxy in this system. Given the well-constrained $z_{\rm spec}$, we fix the redshift to 3.1591 to eliminate potential photometric uncertainties that could affect the modeling of other parameters. The settings for all the other parameters remain the same. The fitting results are summarized in the first column of Table \ref{tab:res}. To allow more flexibility in SFH and to test whether the results are sensitive to the SFH prior settings, we repeat all the analyses with the stochastic prior \citep{2024MNRAS.532.4002W}, which allows the relative importance of different timescales to be quantified. The stochastic prior recovers key galaxy parameters (e.g., stellar mass, metallicity) with the same fidelity as the continuity prior, while its realistic variability enables more accurate and precise fits to galaxy star formation histories. The fitting results based on the stochastic prior are shown in Table \ref{tab:res} as a comparison to the continuity prior.

\begingroup
\renewcommand{\arraystretch}{3}
\begin{table*}
\label{tab:res}
\vspace{0.5cm}
\centering
\begin{tabular}{lccc|ccc}
\hline
 &  \multicolumn{3}{c|}{\large{Continuity Prior}} & \multicolumn{3}{c}{\large{Stochastic Prior}}  \\
\hline
 & Total galaxy & Bar region & Off-bar region & Total galaxy & Bar region & Off-bar region \\
\hline
\makecell[l]{$\log(M_*/M_{\odot})$}       & $10.63^{+0.13}_{-0.13}$ & $10.13^{+0.08}_{-0.08}$ & $10.46^{+0.13}_{-0.14}$  & $10.66^{+0.09}_{-0.09}$ & $10.14^{+0.08}_{-0.09}$ & $10.50^{+0.06}_{-0.09}$ \\
\makecell[l]{$A_V$}                       & $1.13^{+0.08}_{-0.11}$  & $1.19^{+0.33}_{-0.34}$  & $1.12^{+0.14}_{-0.15}$   & $1.10^{+0.11}_{-0.11}$  & $1.34^{+0.32}_{-0.37}$  & $0.78^{+0.24}_{-0.28}$ \\
\makecell[l]{$\log(Z/Z_{\odot})$}         & $-0.54^{+0.16}_{-0.14}$ & $-0.33^{+0.38}_{-0.40}$ & $-0.58^{+0.28}_{-0.26}$  & $-0.55^{+0.15}_{-0.15}$ & $-0.58^{+0.46}_{-0.34}$ & $-0.28^{+0.37}_{-0.45}$ \\
\makecell[l]{$\log U$}                    & $-3.00^{+0.19}_{-0.22}$ & $-4.27^{+1.64}_{-1.08}$ & $-3.55^{+0.34}_{-0.75}$  & $-3.00^{+0.18}_{-0.21}$ & $-4.41^{+1.00}_{-0.80}$ & $-3.13^{+1.25}_{-1.09}$ \\
\makecell[l]{mass-weighted \\ age (Gyr)}     & $0.91\pm0.16$           & $1.03\pm0.11$           & $0.83\pm0.14$     & $0.84\pm0.12$           & $1.00\pm0.10$           & $0.69\pm0.12$        \\
\makecell[l]{recent SFR \\ ($M_{\odot}/yr$)} & $41.55\pm28.98$      & $3.37\pm2.53$       & $36.97\pm24.43$ & $51.51\pm33.06$      & $3.61\pm2.62$       & $48.40\pm15.71$ \\
\makecell[l]{reduced $\chi^2$}                    & 9.30 & 9.76 & 12.48  & 26.42 & 25.04 & 28.32 \\

\hline\hline
\vspace{0.3cm}
\end{tabular}
\caption{Spatially-resolved SED fitting results from \texttt{Prospector} for the entire galaxy, bar region, and the off-bar region. Results from two different SFH prior assumptions are shown side by side.}
\end{table*}
\endgroup
\vspace{1cm}

\begin{figure*}[ht!]
\gridline{\fig{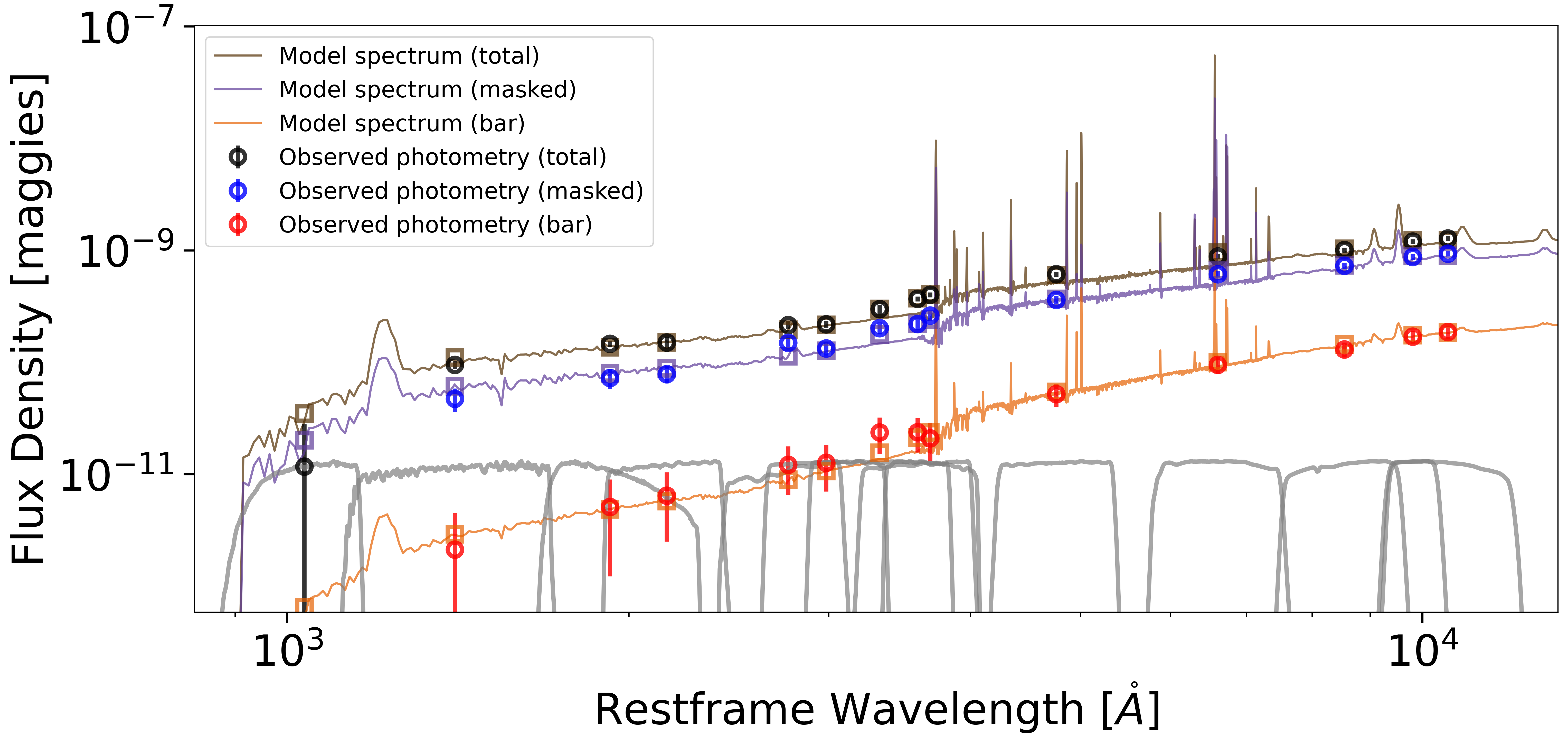}{0.9\textwidth}{(a)}}
\gridline{\fig{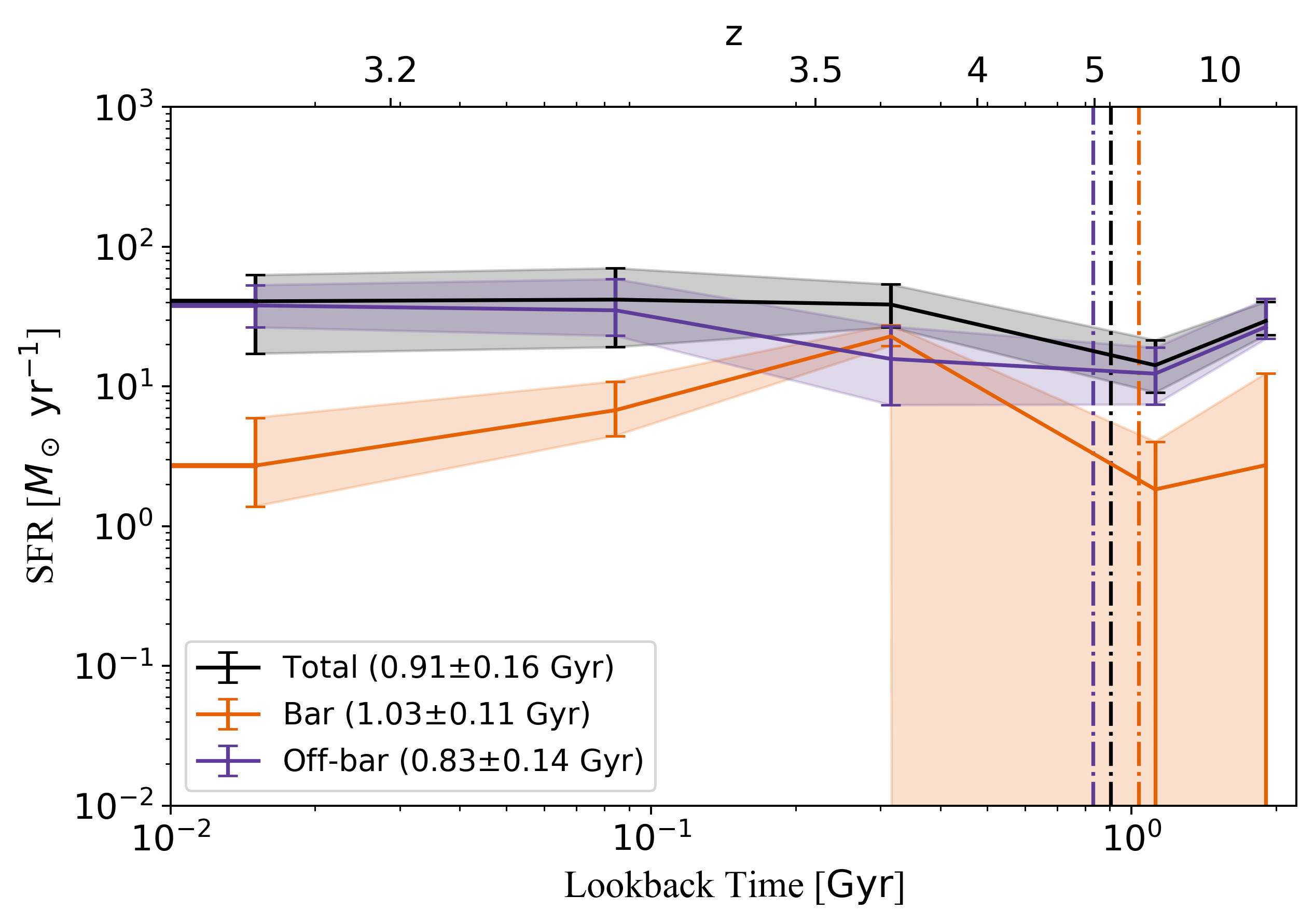}{0.5\textwidth}{(b)}
          \fig{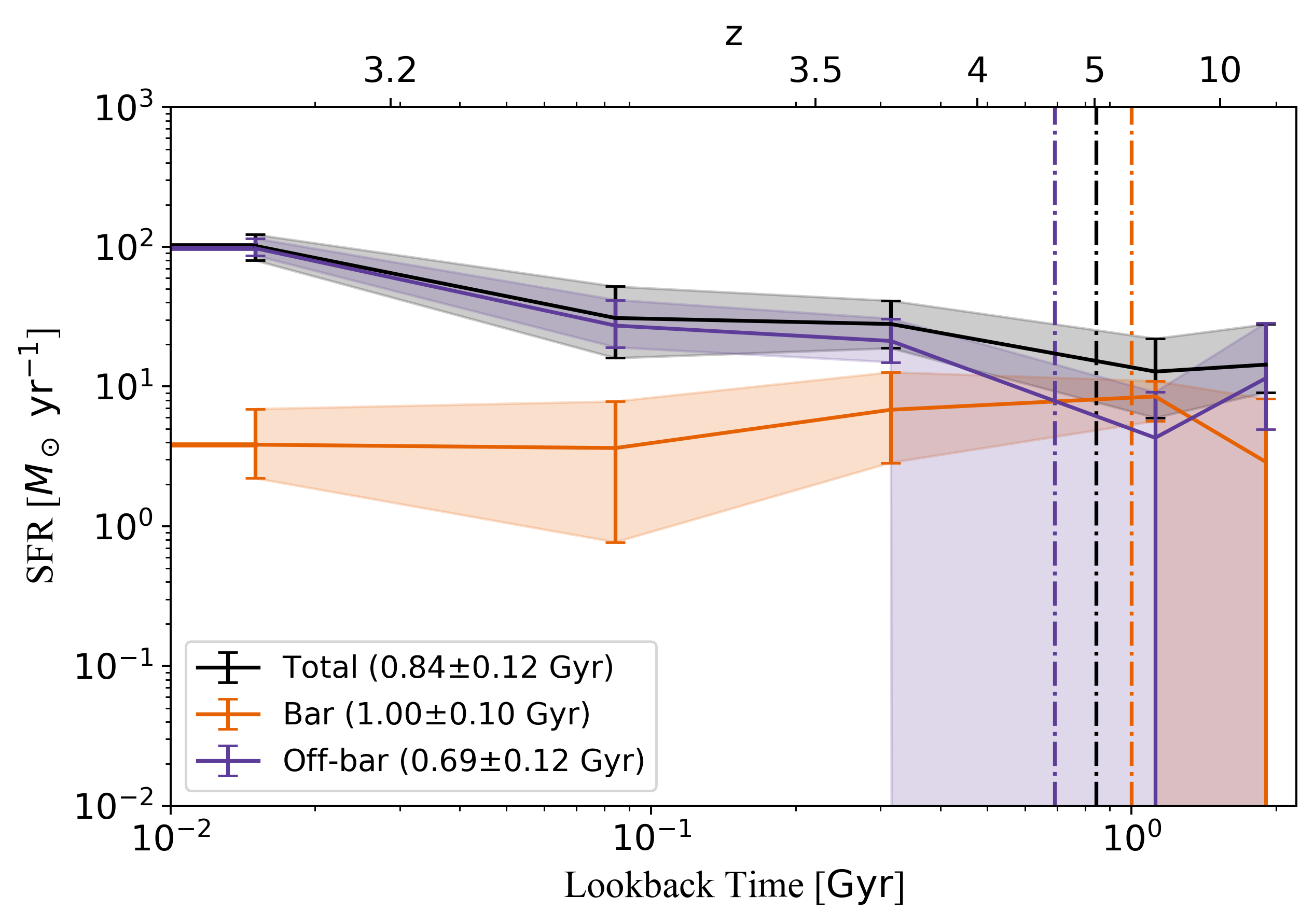}{0.5\textwidth}{(c)}}
\caption{The best-fit spatially resolved SEDs and SFHs of COSMOS-74706. (a) observed photometry data and model spectrum for the original image (black), masked images (blue), and the bar region only (red). The gray lines show the transmission curves of all the available filters. (b) The median 5-bin SFH of the entire galaxy (black), the bar region (orange), and the off-bar region (blue) assuming the continuity SFH prior. The shaded region of each curve shows the uncertainties based on the 16th and 84th percentiles. The mass-weighted ages are marked as vertical dotted lines and labeled in the legend. (c) The same as panel (b) but assuming the stochastic SFH prior.
\label{fig:sed}}
\end{figure*}

To further investigate the lower red component and its barred-spiral features, we apply the masking defined in Section \ref{sec:mask} to avoid overshining of the upper blue component. Using a wide redshift sampling range (uniform between 2 and 7) and keeping all the other settings unchanged, we obtain $z=2.95^{+0.90}_{-0.16}$ for the masked images. This is in agreement with $z_{\rm spec}$ within the uncertainty range, and further rejects the assumption that the red component belongs to another high-$z$ background galaxy. It is more likely that both the red and blue components belong to the same galaxy at $z=3.1591$, or they are two galaxies close to each other, which may be simply overlapping or interacting.
As a next step, we conduct spatially resolved SED modeling for regions both on and off the stellar bar. The bar region is defined by visual identification, as outlined in a white ellipse in Fig. \ref{fig:rgb}. We measure the fluxes of the identified bar structure by summing up the pixel values within the defined region for each band. The fluxes outside of the bar structure are then determined by subtracting the on-bar fluxes from the total fluxes after masking. Given the photometry of both the bar region and the off-bar region, we perform SED modeling again with \texttt{Prospector} to obtain their physical properties and reconstruct the corresponding SFHs. The fitting results are summarized in the second and third columns of Table \ref{tab:res}. We compare the overall and local SFHs with two different SFH prior settings in Fig. \ref{fig:sed}. Similar to the fitting of the entire galaxy, we repeat the analysis with the stochastic SFH prior and compare the results with those of the continuity prior.

\subsection{Explaining the Bimodal Structure} \label{sec:bimodal}

The presence of two distinct bulge-like components with extreme color gradients is scientifically interesting. The bimodal structure can be explained by overlapping galaxies, merging or interacting galaxies, or spatial variations within a single galaxy. So far, no confirmed emission lines associated with a second galaxy have been detected, and the photometric redshift constrained from the masked image remains at $z\sim$3. To further rule out the possibility of the system being comprised of overlapping galaxies, we derive the photometric redshift of the red component only. By applying a small circular aperture around the red bulge-like structure, the short-wavelength contamination from the surrounding structures can be minimized. The SED modeling returns $z=3.04^{+0.14}_{-0.10}$ for the bulge-like structure, which is highly consistent with $z_{\rm spec}=3.1591$. When forcing the redshift prior to be uniform between 4 and 7 to search for any possible high-$z$ solutions, the redshift sampling consistently hits the lower boundary and cannot reach a good solution. Therefore, it is safe to conclude that both the red and the blue components are at a similar redshift.

\section{Results} \label{sec:res}

After masking the upper blue component, the SED modeling returns a photometric redshift $\sim$3, strongly supporting the assumption that this system is composed of a single galaxy. However, more spectroscopic data is needed to completely rule out the possibility that the red component belongs to another galaxy at a similar redshift. In contrast, the presence of a high-z background galaxy at $z>5$ is unlikely. The total stellar mass of this galaxy is constrained as $\log(M_*/M_{\odot})=10.63_{-0.13}^{+0.13}$, suggesting a massive stellar disk. At this redshift, the Universe was only around two billion years old. Despite the large modeling uncertainty arising from its bimodal structure, this galaxy is considered massive at such an early time. This requires a high baryon fraction and high star-forming efficiency at the early evolution stage, yet with no tension with the current $\Lambda$CDM model. 
This galaxy also features high dust attenuation ($A_V\sim$ 1.13) and low metallicity ($\sim$0.3 solar). It is rich in gas and is undergoing active star formation with SFR$\sim$40-50$M_{\odot}/yr$ at the time of observation. According to the total stellar mass and the recent SFR, the galaxy falls on the star-forming main sequence (MS) at $z\sim$3 \citep{2020ApJ...899...58L, 2023ApJ...942...49C}.

The spatially resolved SED modeling enables a comparison of galaxy properties within and outside the identified bar region. The stellar mass within the bar (including the bulge-like structure) accounts for 31.6\% of the total stellar mass, whereas the recent SFR only accounts for 8.1\% of the total amount. Since the bar region occupies only a small fraction of the total surface area (see Fig. \ref{fig:rgb}), it tends to be more concentrated in mass than the rest of the galaxy. The bar region shows a marginally higher dust attenuation, approximately 5-20\% (depending on the choice of SFH prior) higher than the overall attenuation of the entire galaxy. The difference in dust attenuation within and outside the bar is small with the continuity SFH prior ($<1\sigma$), but becomes more significant when adopting the stochastic prior (1-2$\sigma$). As shown in Table \ref{tab:res}, for all the other physical parameters, both SFH priors give comparable results, though the reduced $\chi^2$ of the fit with the stochastic prior is larger than that with the continuity prior. Given the wide posterior distributions, no significant differences are observed in the metallicity or ionization parameter within or outside the bar. 

Most interestingly, the mass-weighted age of the bar region reaches $\sim$1 Gyr, which is 24\% higher than that of the off-bar region with the continuity prior, or 45\% higher with the stochastic prior. In summary, the bar region of this galaxy is more mass-concentrated, more dust-obscured, and less active in forming stars at the time of observation. It is very likely that the bar and the inner part of the galaxy experience bursts of star formation first and then undergo a suppression in star formation due to gas depletion or enhanced turbulence. While the spatial variation of SFH is mildly model-dependent, as shown by the upper panel of Fig. \ref{fig:sed}, the bar region has a distinctly redder SED, aligning with our deduction.

The overall SFH of the entire galaxy is relatively flat, with a tentative rising trend that becomes more pronounced when using the stochastic prior. At early cosmic times, the SFR within the bar region experienced a quick rise within a few hundred Myr, and temporarily dominated the SFR of the whole galaxy. Shortly after this small-scale starburst, the bar region experienced a steady and mild decrease in SFR. In the meantime, the SFR outside the bar gradually increased, growing to dominate the overall SFR. At the time of observation, the SFR within the bar remained significantly lower than the SFR outside of the bar. When assuming the continuity prior, the early burst of star formation within the bar appears to be more striking and reaches the peak at a later time. Also, the overall SFH is flatter compared with the clearly rising trend with the stochastic prior. In either case, the bar region features an older mass-weighted stellar age and is more quiescent now compared with the rest of the galaxy. 

\section{Discussion} \label{sec:dis}
\subsection{Bar Formation and Dark Matter Fraction}
In this study, we explore the properties of a candidate barred-spiral galaxy at $z>3$, when the Universe was only around 2 Gyr old. Baryon fraction has a great impact on the disk instability as it regulates the susceptibility to both internal and external perturbations. Simulations have converged on the idea that the disk-to-total mass ratio ($f_{disk}$) is a factor of primary importance in bar formation and evolution (e.g., \citet{2017ApJ...835...80C, 2023ApJ...947...80B}). 
$f_{disk}$ is defined as the ratio of disk mass to total galaxy mass within the radius at which the rotation curve roughly peaks (2.2 $R_{disk}$). The probability of hosting a stellar bar increases with both the total stellar mass and $f_{disk}$. Most importantly, the bar formation timescale drops precipitously for disks with $f_{disk}>$0.3 (e.g., \citet{2012ApJ...750L..41W, 2018MNRAS.477.1451F}). Therefore, forming a stellar bar in a short timescale requires a high $f_{disk}$, i.e., a low dark matter fraction. To be specific, a bar formation time of the order of 2 Gyr requires at least $f_{disk}=$0.4-0.5 (varying with the stellar mass and the gas mass fraction of the galaxy) \citep{2023ApJ...947...80B, 2024ApJ...968...86B}. Turbulent gas-rich disks at high redshifts require marginally lower $f_{disk}$ to form bars.

Typically, in the local universe, bars encompass from a few percent to 40\% of the total stellar mass of a galaxy \citep{2009ApJ...696..411W}, with a median value of 13\% (e.g., \citet{2011MNRAS.415.3308G}). Moreover, bars appear to enclose a larger proportion of stellar mass in more massive galaxies and galaxies with more prominent bulges \citep{2023gbdd.confE..38M}. COSMOS-74706 is a massive galaxy with total stellar mass $\log(M_*/M_{\odot})\approx10.63$ and a potential bulge-like structure revealed by JWST/NIRCam images. Therefore, we estimate the bar-to-total mass ratio as 10-40\% for COSMOS-74706, indicating a bar mass of $9.63<\log(M_{bar}/M_{\odot})<10.23$. Based on our spatially resolved SED modeling, the stellar mass within the bar region (including both the bar and the potential bulge) is around $10^{10.13} M_{\odot}$. Therefore, the stellar mass of the potential bulge ranges widely from 0 to $10^{9.96}M_{\odot}$. Further kinematics measurements are needed to confirm the presence of a classical or pseudo-bulge and further separate the mass contribution of bulge and bar.

According to Figure. 5 of \citealt{2023ApJ...947...80B} and Figure. 18 of \citealt{2024ApJ...968...86B}, a disk-to-total mass ratio of $f_{disk}>$0.4 is required to form a stellar bar by $z\approx3.16$. Considering a gas mass fraction between 0 and 50\%, the upper limit on the dark matter mass of COSMOS-74706 would be $10^{10.8}$ to $10^{11.1}~M_{\odot}$ (1.5 to 3 times of the disk stellar mass). Dark matter halos of this mass are found to be common in galaxies at $z\sim3$ \citep{2003MNRAS.346..565R, 2017MNRAS.464.1633F}.

In the local Universe, dark matter makes up most of the mass of galaxies ($>$80\%), and becomes more dominant in the outer regions of star-forming disks \citep{1996MNRAS.281...27P}. However, at higher redshifts ($z>2$), a large fraction of massive galaxies are found to be strongly baryon-dominated within the disk scale, where baryons condense at the center of dark matter halos \citep{2017Natur.543..397G, 2017ApJ...840...92L}. While the median value of dark matter fraction remains above 50\% across all galactic scales at $z<2$, it drops below 50\% at $z\sim2.5$, and the decreasing trend appears to extend to higher redshifts \citep{2023arXiv230904541S}. Furthermore, in contrast to the local universe, high-$z$ massive disk-like galaxies show a strong negative trend in the dark matter fraction with increasing baryon surface density \citep{2020ApJ...902...98G, 2021ApJ...922..143P}. In other words, at a given halo mass, $f_{disk}$ increases as the baryon surface density increases, providing ideal conditions for triggering the formation of a stellar bar. Having a $<$60\% dark matter fraction at $z>3$ is highly plausible, setting the stage for the quick onset of stable bar features. Therefore, the detection of a stellar bar at $z>3$ serves as indirect evidence of the baryon-dominated nature of early galaxies at an unprecedented redshift, which is consistent with current theories and numerical simulations.

Apart from the dark matter fraction, the growth of bars is also shaped by the gas fraction, the global rotational energy, and feedback processes. By driving strong outflows, AGN feedback is capable of changing the gravitational potential and reducing the amplitude of (or destroying in extreme cases) an existing stellar bar. Conversely, SF feedback that is distributed across the disk does not provide sufficient spatial and temporal coherence to suppress bar formation \citep{2025OJAp....8E..70W}. The prominent bar feature in COSMOS-74706 suggests weak AGN activity around the time of observation, which is also supported by the low AGN fraction from \texttt{Prospector-$\alpha$} SED modeling. Our results are broadly consistent with the evolutionary pathway proposed in \citealt{2024ApJ...968...86B}, which showed that in disks with significant baryonic dominance, turbulent gas can induce strong radial shear flows that lead to an intermittent star-forming bar within $\sim$500 Myr. Moreover, at high gas fractions, the bar can devolve into a central bulge on $\sim$1 Gyr timescales. The presence of a bulge-like structure with a suppressed SFR relative to the rest of the disk aligns with this picture and suggests that the turbulence-driven bar evolution mechanism might have been at work in galaxies of $\sim$2 Gyr old.

\subsection{Significance of Bar Quenching}

Bar quenching is known as a significant internal mechanism that contributes to the suppression of star formation in disk galaxies, while the underlying nature of this process is not fully understood. In theory, the quenching operates through two complementary processes. First, the stellar bar can efficiently drive gas inside the bar's radius towards the central regions through gravitational torques, thereby depleting the cold gas inside the bar's radius and triggering central starbursts (e.g., \citet{2017MNRAS.465.3729S, 2020MNRAS.492.4697N}). Once the short-lived starburst ends, the central region enter a post-starburst quenched phase (e.g., \citet{2005ApJ...630..837J}). In another scenario, the stellar bar induces strong shocks and shear forces in the surrounding interstellar medium, enhancing turbulence and stabilizing the gas against gravitational collapse, and thus suppressing the star formation (e.g., \citet{2016A&A...589A..66H, 2018A&A...609A..60K}). 
Multi-wavelength imaging and spectroscopy have provided strong evidence for both mechanisms in the local universe, where barred galaxies often show reduced star formation rate in the bar region despite the presence of ample cold gas (e.g., \citet{2005ApJ...630..837J, 2019A&A...621L...4G, 2019A&A...630A..88N}). Recently, \citealt{2025arXiv250507925P} showed that bar-driven gas inflows could be important evolutionary drivers for the dominant population of star-forming galaxies at cosmic noon. 
However, bar quenching typically only affects the inner regions of galaxies, and not all barred galaxies exhibit significant suppression of star formation, indicating that the quenching efficiency depends on a variety of factors, including galaxy morphology, gas fraction, and environments \citep{2021A&A...651A.107G}.

Owing to the difficulty in identifying and resolving barred structures at high redshift, investigations of bar-driven quenching in the early universe are rare. In this study, we are able to derive the spatially resolved SFHs across a barred galaxy at an unprecedentedly high spectroscopically confirmed redshift. As shown in Fig. \ref{fig:sed}, the bar and central region experienced a rapid period of star formation at the very beginning and temporarily dominated the overall star formation within the galaxy. After this local starburst, there was a slow and steady suppression of star formation in the bar and central region, while the SFR outside of the bar gradually increased with cosmic time and eventually overtook the bar SFR. At the time of observation, the bar region features a smaller SFR and an older mass-weighted stellar age than the rest of the galaxy. This result supports the inside-out scenario for galaxy formation and provides tentative evidence for bar quenching at its early stage. In the early universe, the bar quenching mechanism can be more complex due to a high gas fraction and an unstable cosmic environment. The efficiency, spatial scale, and timescale of the quenching could be highly uncertain. More spatially resolved observations at $z>2$ are required to be added to this data point and enable statistical studies.

\section{Conclusion} \label{sec:con}

In this study, we report the identification and characterization of a massive, barred spiral galaxy at $z_{\rm spec}=$3.1591, leveraging high-resolution (sub-kpc scales) HST and JWST imaging combined with spatially resolved SED modeling. Our main findings are as follows:

\begin{itemize}

\item COSMOS-74706 is a massive barred-spiral galaxy comprised of two distinct components with divergent colors. The morphology is highly asymmetrical, with a blue clump in the upper region and a red bulge-like structure surrounded by spiral features in the lower region. 

\item The system is spectroscopically confirmed at $z=$3.1591 with high confidence. Our analysis supports the idea that both components belong to a single galaxy, yet more spectroscopic data is required to rule out the possibility that it is composed of two different galaxies that closely overlap or interact with each other.

\item COSMOS-74706 is a massive star-forming galaxy with a total stellar mass of $\log(M_*/M_{\odot})=10.63_{-0.13}^{+0.12}$ and a recent star formation rate of $\textrm{SFR}\! \approx \,$40-50$\, M_{\odot}/\textrm{yr}$, sitting on the star formation main sequence at its given redshift.

\item According to spatially decomposed SED modeling, the bar region appears to be more evolved and less active in forming stars, with a mass-weighted stellar age 24-45\% older than that of the off-bar region. This supports the inside-out scenario for galaxy formation and provides tentative evidence for bar-driven suppression of star formation in the early universe.

\item The presence of a well-formed stellar bar at $\sim$2 Gyr after the Big Bang implies a $<$60\% dark matter fraction and weak/mild AGN activity in the inner region of this galaxy, consistent with recent findings of baryon-dominated disks at high redshifts.

\end{itemize}

This work provides strong observational evidence that massive disk galaxies can host mature bar-like structures and spatially varied star formation activities as early as $z>$3. This underlines the rapid and complex onset of disk instability and galaxy assembly in the early Universe. Deep $\textrm{R}\geq 1000$ spectroscopy with JWST/NIRSpec would confirm the structure and dynamics of this system, and spatially resolved analysis of similar high-$z$ barred galaxies will further expand these results.

\begin{acknowledgments}
This work is based in part on observations made with the NASA/ESA/CSA James Webb Space Telescope. The data were obtained from the Mikulski Archive for Space Telescopes at the Space Telescope Science Institute, which is operated by the Association of Universities for Research in Astronomy, Inc., under NASA contract NAS 5-03127 for JWST. These observations are associated with program \#1837.

Support for program \#1837 was provided by NASA through a grant from the Space Telescope Science Institute, which is operated by the Association of Universities for Research in Astronomy, Inc., under NASA contract NAS 5-03127.
This research made use of data that were obtained at the W. M. Keck Observatory, operated jointly by the California Institute of Technology, the University of California, and the National Aeronautics and Space Administration. The Observatory's work is made possible by generous financial support from the W. M. Keck Foundation.

J.R.W. acknowledges that support for this work was provided by The Brinson Foundation through a Brinson Prize Fellowship grant.
\end{acknowledgments}

\vspace{5mm}
\facilities{HST(ACS and WFC3), JWST(NIRCam), JWST(MIRI)}

\software{astropy \citep{2013A&A...558A..33A,2018AJ....156..123A}, 
          GALFIT \citep{2010AJ....139.2097P},
          Prospector \citep{2021ApJS..254...22J},
          Python-FSPS \citep{2023zndo..10026684J},
          dynesty \citep{2020MNRAS.493.3132S},
          aperpy \citep{2023zndo...8339191W}
          }

\appendix
\section{Uncertainties in SED modeling}

\begin{figure*}[ht!]
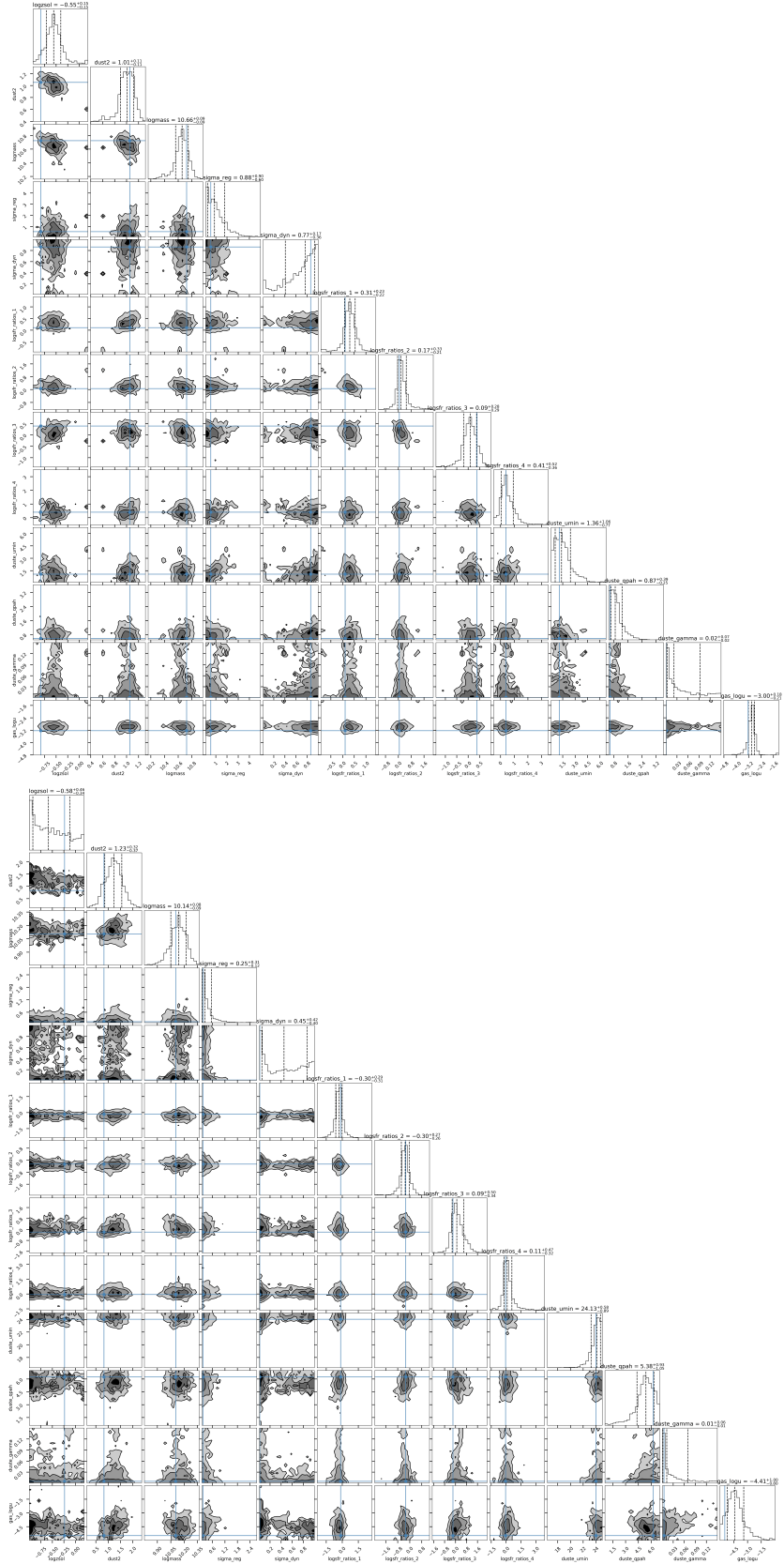

\epsscale{0.72}
\plotone{corner_tot.png}
\plotone{corner_bar.png}
\caption{The corner plots from \texttt{Prospector} SED modeling for the entire galaxy (upper panel) and the bar region (lower panel). \label{fig:corner}
}
\end{figure*}

Throughout the paper, we adopt \texttt{Prospector} SED fitting code with dynamical nested sampling. To demonstrate the fitting uncertainties, we show the corner plots for both the entire galaxy and the defined bar region in Fig. \ref{fig:corner}.

The extended morphology suggests low AGN contribution in this system. To further test the possible existence of an AGN in the bulge-like structure, we run the \texttt{Prospector-$\alpha$}  model on the masked images with an added AGN component. The best-fit AGN fraction is less than 0.01, with an 84th percentile of 0.02. Therefore, we do not consider the potential AGN contribution in SED modeling afterwards.

\bibliography{barref}{}
\bibliographystyle{aasjournal}

\end{document}